# Contemporary Agent Technology: LLM-Driven Advancements vs Classic Multi-Agent Systems

**COSTIN BĂDICĂ[1], AMELIA BĂDICĂ[2], MARIA GANZHA[3], MIRJANA IVANOVIĆ[4], MARCIN PAPRZYCKI[5], DAN SELIŞTEANU[1], ZOFIA WRONA[3]**

[1]*Faculty of Automation, Computers and Electronics, University of Craiova, Bvd. Decebal 107, 200440 Craiova, Romania*
[2]*Faculty of Economics and Business Administration, University of Craiova, Str. A.I.Cuza 13, 200585 Craiova, Romania*
[3]*Faculty of Mathematics and Information Science, Warsaw University of Technology, 00-662 Warsaw, Poland*
[4]*Faculty of Sciences, University of Novi Sad, Trg Dositeja Obradovića 4, 21000 Novi Sad, Serbia*
[5]*Systems Research Institute, Polish Academy of Sciences, 01-447 Warsaw, Poland*

**Abstract.** This contribution provides our comprehensive reflection on the contemporary agent technology, with a particular focus on the advancements driven by Large Language Models (LLM) vs classic Multi-Agent Systems (MAS). It delves into the models, approaches, and characteristics that define these new systems. The paper emphasizes the critical analysis of how the recent developments relate to the foundational MAS, as articulated in the core academic literature. Finally, it identifies key challenges and promising future directions in this rapidly evolving domain.

**Keywords.** Agent Technology, Multi-Agent System, Large Language Model, Agentic AI.

## 1. Introduction

The field of intelligent software agents and multi-agent systems (MAS) has a rich history, with its theoretical foundations grounded in Game Theory and practical achievements manifested through long-standing systems like Jason and JADE and more recent ones including SPADE and MESA. These findings have been enabling the development of agents capable of autonomous action and interaction, often rooted in symbolic Artificial Intelligence (AI) and predefined behavioural rules [1,2].

However, the recent advent of Large Language Models (LLMs) has acted as a profound catalyst, initiating a paradigm shift from symbolic to sub-symbolic AI by dramatically reshaping the landscape of agent and MAS technologies. What was once often considered a niche or "exotic" area of computer science is now experiencing unprecedented traction and demonstrating significant potential for real-world applications.

This transformation is largely due to the LLMs evolving beyond mere sophisticated language processors to become the "cognitive core" or "brain" of a new generation of agents. This evolution signifies a move from traditional AI systems, which primarily responded to user inputs based on explicit programming and manually crafted knowledge bases, to LLM-powered agents that are claimed to actively perceive their environments, reason about goals, and execute actions through continuous learning, dynamic reasoning, and adaptation. Modern LLM agents are simultaneously making leaps forward in multiple areas: (1) they are capable of using a wider range of knowledge sources, (2) have greater capabilities of applying their possessed knowledge to new tasks, and (3) can multimodally interact in more ways than ever before using text, image, and sound [3]. The field has witnessed an explosive growth in research and development, with LLM-based MAS moving from theoretical curiosity to semi-practical reality across diverse sectors [4]. This change isn't just a design choice, but a step to explore ways to overcome the natural limits of a single LLM, when it has to handle complex, dynamic tasks with multiple steps.

The emergence of new LLM-agent designs is fuelling rapid innovation, much like the "Cambrian Explosion" [65]. The danger of this fast growth, however, is that it may encourage a shallow understanding (or even a complete abandonment) of the decades of established principles from the field of MAS. There is a discernible need to ensure that the current innovation is built upon, rather than in ignorance of, the foundational knowledge accumulated in the field, so that efforts are not wasted "revisiting problems the MAS literature has already addressed" [5] and "reinventing the wheel" [6].

Here, note that the very definition of an "agent" is undergoing an implicit renegotiation. Traditional intelligent software agents operate based on explicitly programmed beliefs, goals, and plans. In contrast, LLM-agents derive complex behaviours from the LLM's vast pre-trained knowledge, refined through prompt engineering, memory architectures, external context injection and tool utilization. This knowledge is, however, a big black box, and there is no easy way to understand what is "inside it". This fundamental difference in how agent behaviour is specified and emerges necessitates a careful examination of how classic concepts of agency, intentionality, and autonomy apply in this new LLM-driven context.

This emergent and rapid expansion of LLM-powered agent designs triggered a new surge in terminology of "agentic" and "multi-agentic" AI. We share the opinion of well-known, respected MAS researchers that essentially these terms are just new buzzwords for the long-established field of intelligent agents and MAS with the primary difference being one of historical context, new technological reality and modern branding, rather than establishment of fundamentally novel ideas [6].

The rich history of MAS has provided a robust theoretical framework with practical developments and applications, driven by the seminal works of scholars such as Michael Wooldridge, Yoav Shoham, Katia Sycara, Nicholas R. Jennings, Victor Lesser, Jeffrey S. Rosenschein, Pattie Maes and many others, laying down the fundamental concepts of agent rationality, communication, coordination, cooperation, strategic behaviour, automated negotiation, personal agents, norm emergence and social organization. Agents and MAS is a well-known and very active research area of computer science, with reference events and high-level publication venues: International Foundation for Autonomous Agents and Multiagent Systems (IFAAMAS), Journal of Autonomous Agents and Multi-Agent Systems (JAAMAS), and International Conference on Autonomous Agents and Multiagent Systems (AAMAS). Moreover, research and practice of MAS brought about development of multitude of tools (platforms) [52], even if some of them have not kept up with the current software engineering standards and processes [53].

LLMs meet both game-theoretical foundations [7] and practical developments of MAS [4]. The primary objective of this reflection paper is to critically examine how contemporary LLM agent technologies align with, extend, or potentially challenge the classical approaches. This endeavour is particularly pertinent given the concerns raised within the MAS community itself. Some prominent researchers have cautioned that much of the current literature on LLM-based MAS appropriates established terminology without an in-depth understanding of its foundational principles. There is a potential risk that many systems labelled as "MAS LLMs" may, upon closer inspection, lack the cores of true MAS, such as genuine autonomy and sophisticated social interaction, instead relying on oversimplified, LLM-centric architectures [5]. This paper seeks to explore this tension, providing a balanced perspective on both the transformative potential of LLMs and the enduring relevance of classic MAS theory and practice.

The paper provides our comprehensive reflection on contemporary agent technology, with a particular focus on the advancements driven by LLMs vs classic MAS. It delves into the models, approaches, and characteristics that define these new systems. The paper emphasizes the critical analysis of how these modern developments relate to the foundational approaches of MAS, as articulated in standard academic literature. Finally, it identifies key challenges and promising future directions in this rapidly evolving domain.

The remaining part of this contribution is organized as follows. Section 2 details the architectural models of contemporary intelligent agents and MAS based on LLMs. Section 3 undertakes a focused analysis, reconciling LLM-based agent phenomena with classic MAS theory. Section 4 contains concluding remarks, limitations of our study, challenges and future research frontiers.

## 2. Architectural Models of Contemporary LLM-Agents

The construction of contemporary LLM-based intelligent agents involves a sophisticated interplay of defining their core characteristics, endowing them with memory and planning capabilities, enabling them to act, and structuring their collaboration in MAS. The design philosophy follows a "scaffolding" approach, where the potent, general-purpose LLM core is augmented with specialized modules for functions like memory, reasoning, planning, and tool-based action execution. This hybrid strategy combines the implicit, vast knowledge of LLMs with the precision and control offered by explicit architectural components, echoing some principles of traditional AI agent design but with the LLM now firmly at the operational centre.

### 2.1. LLMs as Knowledge-Based Systems

Before the rise of the intelligent agent paradigm in the 1990s, Knowledge-Based Systems (KBS) were the central and most successful methodology in symbolic AI, particularly during the boom of "expert systems" in the 1980s. This approach was revolutionary for its time, shifting the focus from general-purpose search algorithms to systems that could solve complex problems within a specific, narrow domain. The core innovation of a KBS was the separation of its two main components:
- *Knowledge base* (KB), containing explicit facts and "IF-THEN" rules meticulously gathered from human experts using techniques of knowledge elicitation.
- *Inference engine*, which applied computational reasoning to that knowledge for answering to user-provided queries.

Systems like MYCIN, which diagnosed blood infections, demonstrated for the first time that AI could perform at or above human expert levels on practical, real-world tasks, firmly establishing knowledge representation and reasoning as the dominant focus of mainstream AI research before the community moved towards more integrated agent-oriented models of perception, action, and autonomous goal-seeking.

Over time, these foundational rule-based systems became more formalized, evolving into logic-based systems and deductive databases for more rigorous reasoning. This trajectory continued towards the rich, formal structures of ontologies backed by description logics. They define a common vocabulary and reasoning framework for formally capturing and sharing concepts and their relationships. This approach has culminated in the large-scale, interconnected knowledge graphs that underpin many contemporary AI applications and internet services.

In traditional KBSs, the "Ask/Tell" model is the primary way an agent or human can interact with its KB.
- *Tell service*. This is the mechanism for adding new symbolic statements (rules and facts) to the KB. It is an act of "belief revision". "Telling" a KBS something is actually instructing it to update its persistent representation of the world. For example, *TELL(KB, "Rain is predicted for Craiova today.")* determines the system to incorporate this statement into its KB, potentially resolving contradictions and updating related beliefs. Observe that the knowledge captured by the KB of a KBS is explicit and structured.
- *Ask service*: This is the mechanism to query the KB. The system uses its inference engine to reason over the facts it has been "Told" to derive an answer. For example, *ASK(KB, "Should I take an umbrella today?")* would return a dry "Yes" based on the previously "Told" fact and a rule like *IF rain is predicted, THEN take an umbrella*.

This model creates a KBS whose KB is auditable, editable, explicit and symbolic, while its inference mechanism is deterministic (and, therefore, explainable). A KBS is a "Glass Box" for its developers, sometimes coined "knowledge engineers", as they can trace and check its reasoning step-by-step. However, its main weakness is that it is brittle. If a user asks something that falls outside its pre-programmed rules, it simply fails or gives an "I don't know" response. In theory, this weakness can by tackled by the explicitly expandable KB over time using knowledge acquisition. Nevertheless, this largely manual process has severe practical limitations, failing to support a robust and scalable engineering approach. The primary obstacle that has consistently restricted the broader, practical utility of traditional KBSs is the "knowledge acquisition bottleneck".

As concerning the "Ask" service, from the outside, the user experience with an LLM and KBS can feel similar. The user "Asks" a question, and it provides him or her a response. However, describing a LLM as just a more advanced KBS misses the fundamental difference in how they work. LLMs are not just a scaled-up version of traditional KBSs. They represent a completely different architectural and philosophical approach to AI.

An LLM is truly a "Black Box" operating like a complex network of interconnected parts of ideas. This network encodes *implicit knowledge* in the weights of its billions of neural connections. This "knowledge" is formed by identifying statistical patterns in the vast amounts of training data. Its behaviour leverages on a *generative engine* that stochastically predicts the most probable sequence of tokens that would follow a given prompt, based on the patterns seen and distilled so far.

Perhaps the most significant architectural departure of LLMs and KBs is that LLMs do not have a proper "Tell" service in the classic KBS sense. An LLM's knowledge is implicit and encoded in the weights of its neural network during a massive training event. An LLM cannot be simply "Told" a new fact and have it permanently and reliably revise its core "beliefs". For example, if a "Tell" prompt such as "The new CEO of Google is Costin Bădică" is submitted to an LLM, it does not update a fact in a database, as there is no database to be updated. The model's underlying representation of who the Google's CEO is, remains unchanged.

Instead of a true KBS-like "Tell" service, LLMs have workarounds:

- *In-context learning*: It is a *prompt engineering* technique [51] involving the provision of a fact within the prompt itself (e.g., "Given that the new CEO of Google is Costin Bădică, write him a welcome email."). The LLM will use this "Told" fact during the current query or user session. The next user who asks about the CEO won't benefit from this information.
- *Fine-tuning*: This is the closest equivalent to belief revision, but it is incredibly inefficient. It involves retraining the entire model (or significant parts of it) on new information. It is a costly, complex process that fundamentally alters the LLM.

Because there is no true "Tell", the internal mechanism of an "Ask" service in an LLM is also different from KBS. When a user "Asks" an LLM a question, it is not performing logical inference on a set of explicitly "Told" facts. It is performing probabilistic pattern-matching, generating a plausible sequence of tokens based on the user-provided prompt (including any temporary, in-context information). Here "plausible" should be understood in a wider sense than just stochastic, including also the aspect of "hallucination", i.e. answers that look superficially true, but in fact are false, containing wrong or even fabricated information.

## 2.2. Single LLM-Agent Construction

The creation of an individual LLM-agent is based on several key architectural pillars that collectively define its identity, its ability to learn and remember, its capacity to strategize and perceive the environment, and its power to execute actions and communicate with others (see Fig.1).

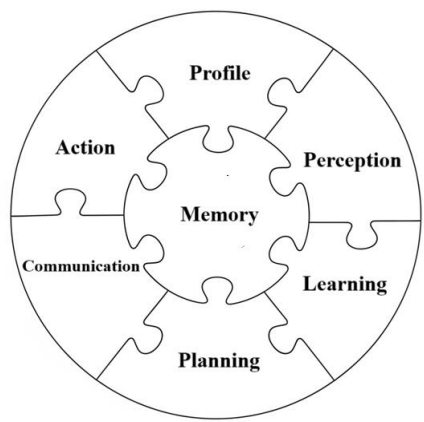

Fig.1 LLM-based agent architectural pillars

**Profile** assumes establishing an agent's operational identity and is a foundational step, achieved by configuring its intrinsic attributes and behavioural patterns. This "profile" can range from simple role descriptions to complex persona definitions that guide the LLM's responses and actions. Methodologies for profile definition include human-curated static profiles, which are often employed in systems like CAMEL [9], AutoGen [10], MetaGPT [11], and ChatDev IDE [12] to ensure domain-specific consistency and adherence to predefined roles, particularly in applications requiring high interpretability or regulatory compliance. Alternatively, batch-generated dynamic profiles, as seen in systems like Generative Agents [13] and RecAgent [14], leverage parameterized initialization to create diverse agent populations capable of emulating complex human societal behaviours, crucial for realistic simulations [3]. These profiles are critical for enabling agents to perform specialized tasks, engage in role-playing scenarios, and exhibit consistent and believable behaviour.

**Memory** enables an agent to exhibit intelligent behaviour beyond immediate stimulus-response. Memory allows agents to retain context from ongoing interactions, learn from past experiences, and accumulate knowledge over extended periods. Contemporary LLM-agent architectures incorporate various forms of memory:

- *Short-term memory* typically involves leveraging the LLM's context window to maintain dialogue history and immediate environmental feedback. Frameworks like ReAct [15] and ChatDev [12] utilize this for transient contextual data endowing agents with contextual reactivity.
- *Long-term memory* overcomes the limitations of finite context windows by employing various strategies to store and retrieve information over longer timescales. These include the creation of skill libraries (e.g., Voyager [16], GITM [17]), repositories of past experiences (e.g., ExpeL [18], Reflexion [19]), and frameworks for synthesizing new tools or capabilities from past learning (e.g., TPTU [20], OpenAgents [21]). Systems like Letta[1] (formerly MemGPT) [22] propose tiered memory architectures to manage different types of information effectively.
- *Knowledge retrieval as memory* goes beyond simple short or long-term storing and retrieving facts. Retrieval Augmented Generation (RAG) has become a prominent technique, allowing agents to access and incorporate information from external KBs, such as text corpora or structured knowledge graphs, effectively expanding their accessible knowledge far beyond their training data.

**Planning** is central to an agent's capacity to define complex tasks and achieve goals. LLMs have demonstrated a strong aptitude for breaking down problems and generating detailed plans [8]. Planning in LLM-agents often involves [3]:

- *Task decomposition strategies* ranging from simpler single-path chaining approaches, where a problem is broken into a linear sequence of sub-tasks (e.g., the "plan-and-solve" prompting paradigm), to more sophisticated multi-path tree expansion techniques (e.g., Tree-of-Thoughts [23]), which allow for exploration of multiple solution paths and backtracking.
- *Feedback-driven iteration* enabling agents to refine their plans by learning from various forms of feedback, including environmental responses, human guidance, model introspection (self-critique), and input from other collaborating agents. This iterative refinement is crucial for adapting to dynamic situations and improving plan quality over time.

---

[1] Letta (previously MemGPT). https://github.com/letta-ai/letta

Agent plans must ultimately translate into actions that affect its environment.

**Action** execution is supported through several means:
- *Tool utilization* is a critical aspect, enabling agents to interact with the external world beyond text generation. Agents can be equipped with tools for precise calculations, accessing up-to-date information via web searches, executing code, interacting with APIs, and controlling other software or hardware systems. Such tools can enable agents to become more domain-specific and dynamic, facilitating their usage in the context of different applications. The process involves both deciding when a tool is needed and selecting the appropriate tool for the task at hand. Frameworks like GPT4Tools [24] and Toolformer [25] exemplify this capability. Another more generalized, increasingly popular approach involves the usage of Model Context Protocol (MCP) servers, which utilize the MCP open-source standardized protocol for connecting LLM-agents with external "toolboxes". In this context, the MCP can be abstracted as an enabler of tool execution by LLM agents by providing a clean and standard way to access tools [55] (see subsection 3.3 for more details).
- *Physical interaction by action execution* for embodied LLM-agents, such as those found in robotics, involves translating high-level plans into physical movements and manipulations within the real world. This requires the LLM to interface with actuators, robotic control systems and interpret sensory feedback from the physical environment. Systems like DriVLMe [26] are exploring this domain.

**Communication and interaction** of LLM-agents is revolutionized by establishing natural language as a universal medium for coordination. The use of natural language enables unprecedented flexibility in agent interactions and can lead to the emergence of complex, adaptive behaviours. LLMs allow agents to generate and understand human-like text, which serves as a novel and intuitive means of guiding and influencing agent behaviours in real-time, facilitating clearer communication and more effective collaboration among agents, as well as between agents and humans.

While the shift towards natural language as the primary interaction medium offers significant advantages in terms of intuitiveness and flexibility, it also introduces new challenges. Natural language is inherently prone to ambiguity and requires careful grounding to ensure common understanding. The reliance on natural language in LLM-based systems raises questions about how these agents achieve robust mutual understanding and avoid the pitfalls of misinterpretation that formal languages sought to mitigate.

**Perception** of LLM-agents is rapidly expanding beyond text. There is a clear trend in the development of foundation models towards native support for multiple modalities, including image generation, video output with improved visual consistency, enhanced voice interaction modes, and the ability to process file attachments such as PDFs and images for richer multimedia analysis. This move towards multimodality is crucial for enabling agents to operate in more complex and realistic environments [48, 49].

**Learning, adaptation and evolution** are hallmarks of intelligent agency. Contemporary LLM-agents are designed to actively engage with their environments, not as static entities, but through processes of continuous learning, reasoning, and adaptation. This adaptive capability is fostered through several mechanisms. Agent evolution is a key area of research, with methodologies focusing on autonomous optimization and self-learning, where agents refine their behaviours by integrating experience, knowledge and feedback from external resources.

Reinforcement Learning (RL), particularly Reinforcement Learning from Human Feedback (RLHF), is being actively explored as a method for training and enhancing the reasoning and planning abilities of LLMs, both during pre-training and post-training stages [27]. The concept of "forward alignment" is also gaining traction; this involves proactively shaping an LLM's behaviour during its training phase by incorporating human-value feedback, aiming to construct more robust systems where the foundation models not only continuously learn but also maintain alignment with human intentions and safety requirements over time [28]. These diverse approaches to learning and adaptation highlight the ambition to create LLM-agents that are not only intelligent at a single point in time but can enhance, improve, and adjust their strategies in response to new information and changing environmental conditions.

## 2.3. LLM-Powered MAS

While individual LLM-agents possess enhanced capabilities, many complex problems require the coordinated efforts of multiple agents. LLM-powered MAS are designed to leverage the collective intelligence of several AI agents, each often powered by an LLM, to tackle tasks that would go far beyond the capabilities of any single agent [4].

We can classify LLM-powered MAS from several different points of view, as each perspective highlights a distinct aspect of their design and function. A comprehensive classification framework may consider the system's architecture, the nature of its agents, and the dynamics of their interaction. Note that similar considerations were made for earlier classic MAS architectures.

**Communication topology and control flow** reveal the most fundamental classification, describing the overall architecture of how agents interact and how tasks are managed.
- *Centralized control* assumes a hierarchical model where a central controller or manager agent (which may itself be an LLM-agent) orchestrates the workflow of activities of other agents. The controller typically decomposes tasks, allocates sub-tasks to specialized or worker agents, and integrates their outputs. Examples include frameworks like CAMEL [9] MetaGPT [11], where centralized manager agents oversee and coordinate worker agents, and AutoAct [29] and AutoGPT[2], where meta-agents or server agents orchestrate specialized agents, possible engaged in structured workflows
- *Decentralized control* involves peer-to-peer interaction among agents without a central coordinator either directly or via a shared environment (broadcast model). There is no single point of control, agents self-organize and coordinate their actions through established protocols or emergent communication patterns, while collaboration can be emergent. Frameworks like CAMEL [9] (partially, for certain aspects, such as the creation of agent societies) and AutoGen [10] (in some configurations) support decentralized interactions.
- *Hybrid control* involve elements of both centralized control and decentralized collaboration, allowing for both local, decentralized collaboration within a team and global, centralized coordination. Examples include KnowAgent [30] and AFlow [31].

**Collaboration model or strategy** focuses on the specific protocols and strategies agents use to work together.
- *Cooperative task decomposition* assumes that a primary agent breaks a task into sub-tasks and assigns them to worker agents.
- *Debate and critique* assumes that agents challenge each other to refine outputs. One agent generates a plan, while a "critic" agent finds flaws, leading to a more robust result.
- *Role-playing and social simulation* assumes that agents adopt specific personas or roles (e.g., Product Manager, Engineer) and are guided by the social norms and behaviours of those roles. This is where emergent social dynamics are most likely to occur, but such situations are also encountered in professional bureaucracy ladders.
- *Negotiation and bidding assumes* agents that dynamically negotiate over resources or responsibilities, often seen in game-theoretic scenarios like auctions.
- *Emergent collaboration or swarm intelligence* assumes that collective goals are achieved without direct commands, as agents act based on simple local rules and observation of a shared environment. This approaches particularly address the study of emergent social dynamics within populations of LLM-agents. Studies have shown that groups of LLM agents can spontaneously develop shared social conventions, such as naming conventions in a coordination game, through interaction alone, without any centralized coordination or explicit programming of these norms. These experiments, often involving agents with limited memory of their recent interactions, mimic the bottom-up formation of norms in human cultures.

---

[2] AutoGPT: Build, Deploy, and Run AI Agents. https://github.com/Significant-Gravitas/AutoGPT

**Agent homogeneity** looks at the nature of the individual agents themselves.
- *Homogeneous LLM-powered MAS* assume that all agents in the system are powered by the same LLM (e.g., all agents use GPT-4). This simplifies design and deployment but might limit the system's overall intelligence to the capabilities of that single LLM.
- *Heterogeneous LLM-powered MAS* assume that agents are powered by diverse LLMs, each potentially specialized for different tasks or possessing different strengths (e.g., one agent uses a powerful proprietary LLM for reasoning, another uses a smaller, fine-tuned open-source LLM for specific data processing). This approach aims to leverage the collective intelligence of varied LLMs for enhanced performance.
- *Hybrid LLM-powered MAS* combine LLM-powered agents with traditional, rule-based, Belief-Desire-Intention (BDI), or non-LLM machine learning agents. This allows for a blend of symbolic reasoning and learned intelligence, leveraging the strengths of both paradigms.

## 3. LLM Agents through the Lens of Classic MAS Approaches

A critical aspect of understanding the contemporary agent landscape is to situate LLM-driven advancements within the rich theoretical framework established by decades of MAS research. This section undertakes such a reconciliation, examining core agent properties, prominent agent architectures like BDI, communication paradigms, and social phenomena like cooperation and norm emergence. A recurring theme is the tension between the rapid, often pragmatic, development of "MAS LLMs" and the rigorous definitions of agenthood and sociality from classic MAS theory. Many current systems, while powerful, might be more accurately described as sophisticated distributed LLM applications rather than true MAS in the classical sense [5]. This highlights a need for careful terminology and deeper theoretical engagement to bridge the gap between innovation and foundational understanding.

### 3.1. Revisiting Classic Core Agent Properties

Classic MAS literature research [41] defines agents by characteristics such as:

- *Autonomy*: Agents operate without direct intervention of humans or others, and have control over their own actions and internal state.
- *Reactivity*: Agents perceive their environment and respond in a timely fashion to changes that occur in it.
- *Pro-activeness*: Agents do not simply act in response to their environment; they are able to exhibit goal-directed behaviour by taking the initiative.
- *Social ability*: Agents are capable of interacting with other agents (and possibly humans) via some agent communication language.

When evaluating LLM-agents against these criteria, a nuanced picture emerges. LLM-steered agents can certainly demonstrate reactivity by responding adaptively to environmental data processed through prompts [50]. Their ability to generate plans and pursue objectives suggests a degree of pro-activeness, especially when augmented with planning modules [3]. The capacity of LLMs to understand and generate natural language inherently provides a powerful medium for social ability [4].

However, the nature and depth of these properties in LLM-agents are subjects of ongoing debate. Authors of [5] argue that many current MAS LLM implementations fall short of embodying true multi-agent characteristics. They contend that LLMs are often fine-tuned as single agents designed to respond to user requests, rather than being trained for genuine interaction with other agents. This can lead to poor performance in tasks requiring a nuanced understanding of other agents' beliefs, desires, and intentions. Furthermore, achieving genuine autonomy is complicated by the LLM's reliance on prompts and its potential non-determinism, which can make behaviour less predictable and controllable than in traditionally programmed agents. The assessment of reactivity, proactiveness, and social ability is often best tested when agents are given high-level goals that require them to dynamically perceive, take initiative, and interact to achieve objectives, scenarios which are becoming more common in LLM agent research.

### 3.2. The BDI Model in the Age of LLMs

The BDI model, inspired by Bratman's philosophy of action and practical reasoning [34] and formalized by researchers like Rao and Georgeff, has been a cornerstone of rational agent architecture and programming leading to the AgentSpeak family of agent-oriented programming languages [32] with implementations like Jason [33]. The model assumes that agents maintain:
- *Beliefs*: Information about the state of the world (possibly incorrect or incomplete).
- *Desires*: Motivational states representing objectives to be achieved.
- *Intentions*: Commitments to achieve a subset of desires typically operationalized as plans.
- *Plans*: Sequences of actions to achieve intentions.

The advent of LLMs raises questions about how these mentalistic notions apply. Can an LLM genuinely "believe" or "desire" or is it merely simulating these states through sophisticated pattern matching and text generation based on its training data and current prompt? Here, we acknowledge that the discussion about the existence and nature of mental states of LLMs is ongoing [56] and that its final outcome (if there will be one reached, by the philosophers) is secondary to this work. While, it is relatively clear that LLMs do not possess beliefs or desires in the human sense, the BDI framework is proving to be a valuable conceptual tool for structuring LLM-agent behaviour. Classic agent concepts like BDI are not being discarded but are being re-interpreted and operationalized through the natural language and reasoning capabilities of LLMs. This suggests an evolution where the implementation of these mental attitudes is changing, rather than the foundational concepts themselves becoming obsolete.

The advent of LLMs provides a powerful new dimension to BDI agents, addressing some of their traditional limitations, especially in handling unstructured data and complex reasoning. LLMs can enhance BDI agents in several key ways:

- *Enriched beliefs*: LLMs can digest and "understand" vast amounts of natural language, enabling agents to build more nuanced and comprehensive beliefs about their environment, e.g. interpreting complex sensor data, visual scenes, web content, and conversational inputs.
- *Flexible desires and goals*: LLMs can help to formulate and prioritize desires by interpreting high-level, ambiguous human orders and translating them into concrete, actionable goals for the agent.
- *Dynamic intentions and planning*: Instead of relying solely on pre-programmed plan libraries, LLMs can dynamically generate or refine plans based on current beliefs and desires. They can reason through complex scenarios, break down large tasks into smaller subgoals, and adapt plans as circumstances change.
- *Natural language interaction*: LLMs empower BDI agents to communicate with humans in a more natural and intuitive way, understanding complex instructions, generating human-like responses, and even inferring human mental states.
- *Tool use and external integration*: LLM-powered BDI agents can effectively leverage external tools, APIs, and databases. The LLM acts as a reasoning engine to decide when an external tool is needed, identify the correct tool, formulate inputs, and interpret outputs, significantly expanding the agent's actionable capabilities.

Despite the potential benefits of LLM and BDI fusion, challenges remain:

- *Computational cost and latency*: LLM inference can be computationally intensive and incur latency, which might be a concern for real-time BDI applications.
- *Hallucinations*: LLMs can sometimes generate incorrect or fabricated information, which can impact the accuracy and correctness of an agent's beliefs and decisions. Here, it should be noted that differentiating hallucinations from facts is a research question in its own right [54].
- *Lack of determinism / uncertainty*: The probabilistic nature of LLMs can make it difficult to guarantee deterministic behaviour (the same prompt may result in a different response), posing challenges for critical applications requiring high reliability.
- *Integration complexity*: Combining the symbolic reasoning of BDI with the statistical nature of LLMs requires careful architectural design and engineering.

A notable example of this integration is the NatBDI architecture [34]. It demonstrates how the BDI model can be adapted to allow instruction and programming of agents using natural language, enhancing transparency and potentially reducing extensive retraining when plan modifications are needed. NatBDI leverages the BDI reasoning cycle but adapts its components for natural language environments.

- Beliefs are represented as natural language descriptions derived from perceptions, making the agent's mental state better aligned to humans than dry symbolic literals.
- Plans are encoded by developers using a controlled natural language, forming a plan library.
- Plan selection is achieved using Natural Language Inference (NLI) models to determine if a plan's context (also in natural language) is entailed by the agent's current beliefs.
- A RL component serves as a fallback policy to generate actions when no suitable plan is found through NLI.

In this context, ChatBDI [35] is a novel agent architecture that integrates the strengths of LLMs with the classic BDI framework based on Jason [33]. Essentially, it uses an LLM as the core "reasoning engine" to update the agent's beliefs and select its intentions, while retaining the BDI structure to provide a clear, explainable, and goal-directed "skeleton" for the agent's behaviour. This hybrid approach aims to create more sophisticated and understandable autonomous agents by combining the powerful, flexible reasoning of LLMs with the formally grounded and predictable decision-making process of the BDI model.

Closely related to the BDI model is the concept of Theory of Mind (ToM) firstly introduced by David Premack and Guy Woodruff in 1978 [36]. While there is significant research on computational models of ToM, unlike BDI, which has a well-defined and widely adopted architectural model in MAS, there isn't a single, universally agreed-upon "ToM architecture / framework / programming language" built around in the same way. Instead, researchers build computational ToM models by leveraging existing MAS (LLM-based or classic) tools and paradigms, focusing on the specific aspects of ToM they aim to simulate such as inferring goals, predicting actions, and understanding belief formation.

ToM captures the ability to explain intelligent behaviour by attributing "mental states" (beliefs, desires, intentions, emotions, knowledge) to others and to understand that others have states that are different from one's own. Research is increasingly exploring ToM capabilities in LLMs. The ToM-agent paradigm aims to empower LLM-based generative agents to simulate ToM in open-domain conversations. It achieves this by disentangling the agent's confidence about an inferred mental state from the mental state itself (e.g., belief, desire, intention of a counterpart). Using past conversation history and reflective reasoning, a ToM-agent can dynamically adjust its inferences about a counterpart's BDIs and its confidence in those inferences. Experimental findings suggest that such agents can grasp underlying reasons for a counterpart's behaviour beyond simple semantic understanding or common-sense decision-making [37]. However, other studies indicate that LLMs, particularly without specialized architectures or prompting, still face significant challenges in robust ToM reasoning, especially in inferring others' intentions and needs in complex coordination scenarios [38].

### 3.3. From Agents & Artifacts to LLM-agents with Tools

A foundational principle in MAS engineering is the explicit separation of agents from their environment. This separation is not merely a design choice but a conceptual necessity that allows for modularity, reusability, and a clear demarcation of concerns. An agent's intelligence and autonomy are defined by its internal state and decision-making processes, while the environment represents the external context in which the agent operates, perceives, and acts. In classic MAS, achieving this separation required standardizing the interface between agent and environment, defining precisely how agents could perceive and influence their surroundings. The advent of LLM-agents has revisited this fundamental challenge, replacing formally defined APIs with more fluid, language-driven interactions with external "tools". This section explores two distinct yet related approaches to mediating the agent-environment boundary: the Agents & Artifacts (A&A) meta-model from classic MAS and the contemporary paradigm of LLM-agents with tools, highlighting their conceptual and technological parallels and divergences.

**The A&A meta-model**, introduced by Ricci, Viroli, and Omicini [61], sought to elevate the environment from a passive backdrop to a first-class abstraction in MAS design. The core idea was to reify the environment's resources and services into programmable entities called "artifacts". Instead of interacting with a monolithic, undifferentiated environment, agents would interact with specific, well-defined artifacts.

Conceptually, an artifact is an object that agents can use to achieve their goals. It exposes a set of operations (actions an agent can perform on it) and observable properties (information an agent can perceive from it)—in some sense similarly to a software service. So artifacts are explicitly designed to be stateful, persistent entities within a shared environment defining the agent's "workspace". You can think of an artifact like a shared blackboard or a calendar; its primary function is to hold a state that can be observed and manipulated by multiple agents over time, serving as a coordination mechanism.

For example, in a collaborative writing scenario, a shared document could be modelled as an artifact. Agents (human or artificial) could interact with it through operations like *write(text, position)* or *lockSection(sectionID)* and perceive its state through observable properties like *content()* or *isLocked(sectionID)*. The artifact itself could encapsulate the logic for managing concurrent access, versioning, and enforcing rules, thus offloading this complexity from the agents.

Technologically, this was realized through frameworks like CArtAgO (<u>C</u>ommon <u>Art</u>ifact Infrastructure for <u>Ag</u>ent <u>O</u>pen Environments), which provides an API for creating and managing artifacts. Agents, often implemented in BDI-style languages like AgentSpeak, would use this API to "focus" on an artifact, discover its available operations, and invoke them. The key innovation was the standardization of this interface. The *use* primitive in CArtAgO allowed an agent to perform an action on an artifact, which would, in turn, affect the environment and potentially be perceived by other agents. This created a structured, predictable, and verifiable medium for agent interaction, where the environment's "laws" were encoded within the artifacts themselves. The A&A model provided a powerful abstraction for engineering complex coordination and cooperation patterns, making the environment an active, programmable part of the system.

An implementation of CArtAgO is JaCaMo [62] built on top of Jason framework [33]. JaCaMo is a framework for MAS programming that combines three different technologies: Jason for agent programming, CArtAgO for environment handling via A&A meta-model, and Moise+ for organizational modelling.

**LLM-Agents with tools and the MCP protocol**. The rise of LLM-agents has introduced a new paradigm for agent-environment interaction. Here, the "environment" is the vast digital world of APIs, databases, and external services. An LLM-agent's ability to act beyond generating text is contingent on its capacity to use "tools". A tool, in this context, is analogous to an artifact: an external resource that the agent can use to achieve a goal, such as a web search API, a calculator, a code executor, or a database query interface.

The core challenge remains the same as in classic MAS: how to standardize the interface between the agent and these tools. The dominant approach relies on directly using the LLM's reasoning capabilities. The agent is provided with descriptions of available tools (in natural language) and, through in-context learning or fine-tuning, learns to decide when to use a tool, which tool to use, and what parameters to provide. The interaction is conversational: the LLM generates a specific function call (e.g., a JSON object specifying the tool name and arguments), which is then executed by a hosting framework, with the result being fed back to the LLM as text.

To address the need for a standardized, scalable, and secure way to manage this interaction, protocols like the MCP have been proposed [55]. MCP aims to be a standardized layer between LLM-agents and the "toolboxes" they access. Conceptually, an MCP server acts as a gateway or broker for tools. It exposes a manifest of available tools and their functionalities, allowing an agent to discover what it can do. When an agent decides to use a tool, it sends a request to the MCP server, which then handles the secure execution of the tool and returns the result. MCP can be seen as an "enabler" of tool use, providing a clean separation between the agent's reasoning core and the external environment of tools. It aims to create an open, interoperable ecosystem where any MCP-compliant agent can interact with any MCP-compliant tool server.

While separated by different technological eras, the A&A meta-model and the LLM-with-tools paradigm share a common conceptual goal: structuring the agent-environment interaction. However, their philosophies and implementations reveal critical differences.

Conceptually, both artifacts and tools are first-class environmental abstractions designed to be manipulated by agents. They encapsulate functionality and expose an interface for interaction, effectively mediating the agent's action on the world. There are, however, important conceptual differences. The first difference lies in the nature of this interface and the reasoning required to use it. In A&A, the interface is formal and explicit. An agent's reasoning is based on symbolic planning; it selects a pre-defined plan that involves invoking a known artifact operation. In contrast, the LLM-agent's interface is semantic and implicit. The agent's reasoning is generative and probabilistic; it infers the need for a tool and constructs the appropriate call "on the fly" based on its understanding of the tool's natural language description and the current context. This makes the LLM-agent approach more flexible and adaptable to novel situations but less predictable and verifiable than the A&A approach [5]. The second conceptual difference is that while A&A meta-model treats artifacts as persistent entities with their own identity, observable properties, and defined operations, an LLM agent's tools, however, are generally not modelled as persistent entities. The "environment" for an LLM agent is often just the set of tool definitions (function signatures and descriptions) provided in the prompt, which the LLM uses to decide which function to call. The tool is less of an object in the world and more of a capability the agent can access.

From a technological standpoint, the A&A meta-model's strength is its formal rigor and predictability. By using a standardized API like CArtAgO, developers could create robust and verifiable MAS where agent interactions were governed by the explicit rules of the artifacts. The primary disadvantage was its rigidity and the significant engineering effort required to define ontologies and artifact functionalities.

The LLM-with-tools paradigm, especially when mediated by a protocol like MCP, offers more flexibility and scalability. It lowers the barrier to entry, as developers do not need to define complex formal ontologies; they simply need to provide natural language descriptions of their tools. The LLM handles the "heavy lifting" of reasoning about tool use. This approach excels at handling the unstructured and dynamic nature of the modern digital environment. However, this flexibility comes at the cost of the "three Cs": cost, controllability, and certainty [6]. LLM inference is computationally expensive. The non-deterministic nature of LLMs makes agent behaviour less predictable, and ensuring that an agent will reliably and safely use tools is a significant challenge, opening avenues for prompt injection and other security vulnerabilities.

In essence, viewing LLM tools as artifacts would be a significant oversimplification. Artifacts in the A&A model are rich, stateful, and structural components of a shared world, designed to be explicitly manipulated for coordination. LLM tools are more akin to stateless capabilities that the agent's reasoning engine (the LLM) can dynamically decide to invoke.

Conceptually, some modern LLM-based agent frameworks are adopting the language of the A&A model. For instance, frameworks like CrewAI [64] describe the tools, APIs, and datasets that their LLM agents interact with as "artifacts". This aligns with the A&A principle of viewing the environment not as a passive backdrop, but as a collection of active resources and tools that agents can manipulate to achieve their goals. However, a deeper formal description of artifacts in the sense of A&A is still lacking.

In conclusion, the journey from A&A meta-models to LLM-agents with tools represents a classic AI trade-off: the exchange of formal, verifiable, and rigid systems for flexible, powerful, but less predictable ones. The A&A meta-model provided a principled way to engineer the environment, making it a predictable and programmable component of the MAS. The LLM-with-tools paradigm leverages the emergent reasoning of LLMs to navigate a much larger and more complex environment, with protocols like MCP attempting to impose a necessary layer of structure and security, but a lower level of formal conciseness. Both approaches underscore the enduring importance of separating agent from environment, and the evolution of this separation reflects the broader shift in AI from symbolic, logic-based reasoning to sub-symbolic, generative intelligence.

Moreover, while the classic A&A meta-model, with its emphasis on stateful, object-like artifacts, is different from the often stateless, API-call-based "tools" used by many current LLM agents, the field is moving toward a synthesis. Researchers are exploring how the structured, explicit environmental interactions of the A&A model can bring more robustness and coordination to the dynamic, flexible reasoning of LLM-powered MAS. Nevertheless, far more research work in this direction is expected.

### 3.4. From Agent Communication and Speech Act Theory to LLM-Mediated Dialogue

The challenge of enabling autonomous entities to communicate, coordinate, and negotiate is fundamental to AI. For decades, this problem was addressed through formal, structured protocols. However, the recent advent of LLMs has introduced a paradigm shift, moving from explicit, machine-readable languages to implicit, human-like dialogue. This section traces the evolution of agent communication from its philosophical foundations to the current state of the art.

**Speech Act Theory.** The classical approach to agent communication is rooted in Speech Act Theory (SAT), a philosophical concept introduced by J.L. Austin [39] and refined by John Searle [40]. The theory posits that natural language utterances are not just stating facts but are themselves "speech acts" like actions possibly changing the state of the world. They distinguish three facets of a speech act: i) *locution* that refers to the content of the utterance; ii) *illocution* that refers to the intention of the emitter; iii) *perlocution* that refers to the effect of the utterance on the receiver. For example, the statement as an illocutionary act "The window is open" could be an *inform* act, a *request* to close it, or a *warning* about the cold, depending on the context and intent.

This powerful idea was directly translated into computational frameworks through definition of ACLs. The two most prominent standards were KQML (Knowledge Query and Manipulation Language) and its more formalized successor, FIPA-ACL (Foundation for Intelligent Physical Agents - Agent Communication Language) [41].

The philosophy of ACLs was to make the illocutionary act explicit and unambiguous. A message in FIPA-ACL is a structured object containing several key components, most notably the performative, which explicitly states the message's intent.

For instance, a *BuyerAgent* requesting the price of *item-66A* to a *ShopAgent* would construct the following formal message:

```
(request
  :sender    BuyerAgent
  :receiver  ShopAgent
  :language  SL
  :ontology  retail-ontology
  :content   "price(item-66A,P)"
)
```

The strength of this approach consists in its clarity, predictability, conciseness and efficiency. The performative *request* leaves no doubt about the sender's intent. Communication is machine-readable, verifiable, and reliable, allowing for the formal design and analysis of multi-agent interactions. This allowed fast processing of "small" messages [60]. The question of efficiency of natural language based communication (e.g. message size and speed of message processing) remains open. This is particularly so for large scale deployments with thousands of agents that are aimed to "talk to each other".

However, this rigidity is also its primary weakness. ACLs require all agents to share a pre-defined ontology and adhere to a strict protocol, making them brittle and unable to handle nuance or novel situations gracefully.

**LLM-mediated dialogue.** Instead of relying on a formal, structured language, LLM-powered agents can communicate directly using natural language. The burden of understanding intent shifts from parsing a performative to interpreting the semantic meaning of unstructured text.

In this new paradigm, the previous example would look starkly different by letting *BuyerAgent* to send the following *prompt* to *ShopAgent*:

    *BuyerAgent* to *ShopAgent*: "Find out how much item 66A costs!"

Here, the illocutionary act of a *request* is not explicitly declared via a performative but is implicitly understood by the receiving agent through its general language capabilities. This approach offers profound advantages:

- *Flexibility and adaptability*: Agents do not need to share a rigid ontology. They can negotiate meaning, clarify ambiguity, and adapt their communication strategy on the fly.
- *Expressive power*: Natural language can convey complex nuance, context, and even social cues that are impossible to encode in a formal ACL.
- *Reduced overhead*: It eliminates the need for developers to design and share complex communication protocols and ontologies.

Frameworks like "Sentimental Agents" [42] explore deliberation and opinion dynamics in LLM-based MAS using natural language interactions. These agents are equipped with "Mental Models of Self" and use sentiment analysis to interpret and react to communications. While they simulate linguistic interactions and analyse the sentiment and content of messages, they do not explicitly employ SAT formalisms.

This shift, however, is not without its own challenges. The explicit intent of the ACL performative is replaced by the implicit guidance of prompt engineering. An agent's behaviour, its willingness to cooperate, and its interpretation of a message are now shaped by its initial prompt. Here, recall that the same prompt may, also, result in different responses, leading to system unpredictability. Therefore, the predictability of the formal system is traded for the emergent, and sometimes unpredictable, behaviour of the LLM. For example, from the security perspective, the usage of unconstrained unstructured natural language creates more opportunities to embed malicious content or intents that are challenging to identify and can hinder the trustworthiness and fairness of the communicative acts. Furthermore, issues of ambiguity, potential for misinterpretation, hallucinations, and the computational cost of processing natural language become primary concerns.

Observe that in essence, the field is experiencing a full-spiral moment. We are transiting from the explicit declaration of intent in ACLs to an implicit intent inferred from natural dialogue. While this might appear to be "reinventing the wheel" [6], it is more accurately described as building a new kind of wheel—one that is infinitely more flexible and powerful, but whose mechanics and failure modes are still being actively explored (with no clarity as to the final answers). The foundational concepts of SAT are arguably now more relevant than ever, not as a blueprint for a rigid language, but as a framework for analysing and steering the complex, emergent dialogues of contemporary "multi-agentic" AI.

A major issue of concern is the lack of standardized protocols for LLM agents to communicate with other agents, external tools, data sources, robots and humans. Ad-hoc proposals motivated by specific applications have been recently surveyed and classified along two dimensions into: i) context-oriented versus inter-agent protocols, and ii) general-purpose versus domain-specific protocols [43].

The Agora protocol represents an interesting development in this space [44]. It is proposed as a meta-protocol that leverages LLMs' capabilities to enable agents to adopt and understand various communication protocols based on the context of the interaction. This could potentially bridge the gap between the flexibility of natural language and the precision of more formal communication structures, allowing agents to dynamically choose the most appropriate "language game" for a given situation. However, it is also clear that today (August 2025) proposals like Agora, (i) are in very early stages and (ii) may not be widely acceptable. The latter reminds, those who know the history of MAS, the failed attempt at MAS standardization via the FIPA standards [59].

### 3.5. From Coordination, Cooperation, and Negotiation in MAS to LLM Agent Performance of Emergent Behaviours

The ability of autonomous agents to interact effectively hinges on three fundamental capabilities: coordinating their actions to avoid incoherence and interference, cooperating to achieve shared goals, and negotiating to resolve conflicts of interest. Classic MAS theory offers a wealth of models and mechanisms for these processes. The architectural shift from classic MAS to LLM-powered agents has profoundly redefined the mechanisms behind each of these pillars, moving from explicit, mathematically-grounded interaction protocols to implicit, dialogue-driven emergent strategies.

**Coordination** is the act of managing interdependencies between agents' actions to ensure the system behaves coherently.
- *Classic MAS*. Coordination is typically achieved through explicit mechanisms—interaction protocols and pre-defined social conventions. Agents are endowed with particular behavioural rules and procedures that allow them to manage and share resources and to make sure their actions don't clash. For example, traffic light protocols for physical robots, or standardized scheduling and locking mechanisms for shared data. The FIPA Request-When protocol is a perfect example, providing a formal, unambiguous way for one agent to ask another to notify it when a certain condition is met, thereby coordinating their timing.
- *LLM agents*. Coordination is an emergent property of context-aware dialogue. An LLM agent doesn't need a formal protocol to avoid conflict; it infers the need for it from the conversation and its understanding of the shared environment. If one agent announces, "I will now summarize the meeting notes", another agent will understand not to perform the same task. This implicit deconfliction is incredibly flexible but lacks any formal guarantees of classic systems. Its success depends entirely on the LLMs' shared understanding and the clarity of their communication.

For coordination, a key challenge for LLM agents is creating benchmarks and understanding the innate capabilities of LLMs to synchronize actions. A framework for evaluation is introduced in [38] that tests ToM reasoning and joint planning capabilities of LLM agents to act in coordination scenarios. Research results indicate that LLM-agents can excel in coordination tasks where decision-making primarily relies on understanding environmental variables and game rules. However, they face greater challenges in scenarios that demand active consideration of partners' beliefs, intentions, and hidden information. This suggests that while LLMs possess strong environmental comprehension, their ability to reason strategically about other agents' mental states in complex coordination games is still developing being debated in the research literature [57], [58].

**Cooperation** goes beyond simple deconfliction, involving a proactive effort to work together towards a common goal.

- *Classic MAS*. Cooperation is structured around distributed problem-solving and shared plan execution. Frameworks like the FIPA Contract-Net protocol allow an agent to decompose a task and "outsource" sub-problems to other agents, who then cooperate to solve the larger problem. Agents often maintain explicit models of each other's capabilities and goals to offer help, sharing and merging complex plans to achieve a collective objective.
- *LLM agents*. Cooperation is driven by a shared intent, typically established in a common system prompt. A team of LLM agents might be instantiated with a prompt like, "You are a team of expert researchers collaborating on a report about climate change. Work together to produce the most comprehensive and accurate document possible". This fosters a "cooperative spirit" as a behavioural trait. The agents then should use natural language to brainstorm, divide labour, and merge results, achieving cooperation through fluid conversation rather than by executing a formal, distributed plan. However, experimental support that this will really happen (and will happen every time) is lacking.

Research into cooperation often explores how LLM agents can overcome social dilemmas or act proactively to assist others in achieving a shared goal. The ProAgent framework was designed to build cooperative LLM agents that can proactively assist humans or other agents [45]. The evolution of cooperation in LLM-agent societies was recently studied by simulation of game-theoretic paradigms like the $n$-player Diner's Dilemma. LLM-agents are assigned different strategies (e.g., always cooperate, defect until punished, punish defectors). Preliminary findings are promising: replication of the evolution of cooperation dynamics predicted by traditional mathematical models, effectiveness of explicit punishment mechanisms in driving the emergence of cooperative norms and reinforcing cooperative behaviour, and high accuracy of LLM agents in interpreting their assigned strategies and choosing appropriate actions [46].

**Negotiation** is required when agents have conflicting goals and must find a mutually acceptable compromise or consensus through exchange of offers and counter-offers.

- *Classic MAS*. Negotiation is a formalized process, rooted in game theory and mathematical optimization [41]. Rational agents are designed with explicit utility functions capturing their preferences. They engage in structured negotiation protocols (e.g., English auctions, alternating offers) and use private strategies to maximize their personal utility. Argumentation-based negotiation allows agents to exchange arguments to persuade others to change their beliefs or preferences.
- *LLM agents*. Negotiation becomes a process of persuasive rhetoric and creative problem-solving. An LLM agent doesn't calculate Nash equilibrium; it leverages its vast training on human text to employ tactics like framing, making concessions, finding creative compromises, or appealing to fairness. The "utility function" is often implicitly defined in its prompt (e.g., "You are a savvy negotiator who values long-term relationships over short-term gain"). This makes the negotiation process more human-like and capable of resolving complex, multi-issue problems that are difficult to model mathematically, but it also makes the outcome less predictable and harder to analyse formally.

Recent research on LLM agents' negotiation focuses on moving beyond simple bargaining to more complex, human-like negotiation, often using self-play and structured feedback to improve performance. A new framework where LLM agents engage in bargaining scenarios and receive structured, utility-based feedback on their performance is introduced in [46]. The results show that this feedback method helps the agents think more strategically and understand their opponent's viewpoint, which makes them look like more successful human negotiators. However, the meta-level question remains. Is it really "better" for negotiating parties to be human-like, or should the "human-factor" be removed in order to reach "better" negotiation outcomes (and reach them faster).

### 3.6. Norms and Social Laws in Classic MAS vs. LLM-Agent Societies

For any society of autonomous entities to function without descending into chaos, it requires a framework of rules, norms, and laws. This framework governs interactions, manages shared resources, and ensures that individual actions contribute to a stable and productive collective. The approach to defining and enforcing these rules has undergone a fundamental transformation, evolving from the rigid, logic-based systems of classic MAS to the fluid, natural language-based principles of LLM-agents.

**Classic MAS** articulates explicit, machine-readable, and verifiable norms and social laws as formalized constructs. The primary goal was to create predictable and mathematically verifiable systems. If agents followed the rules, a system designer could formally prove that certain undesirable states (like gridlock or resource depletion) would never occur.

- *Representation*: Norms were typically represented using formal logic, most notably deontic logic, which deals with the concepts of obligation ($O$), permission ($P$), and prohibition ($F$). A social law could be encoded as an unambiguous rule like $O(agent\_A, pay\_fee)$ *IF enters\_park* that can be directly processed by an agent's reasoning engine. The system's entire normative framework was a set of these explicit rules, agreed upon beforehand.
- *Enforcement* was an active, computational process. It was typically handled through two main mechanisms:
    - *Centralized monitors* represented by specialized "guardian" or "norm-enforcer" agents that observe the actions of other agents and have the power to apply sanctions (e.g., fines, exclusion) to those who violated a norm.
    - *Decentralized sanctions* assumed that norms were upheld by the agents themselves. If an agent detected a violation by another, it could apply a sanction directly, creating a self-policing system.

The key ingredient is that in classic MAS, norms are external constraints that are formally defined and computationally enforced, much like the laws of physics in a simulated world.

**LLM-agents** employ more implicit, contextual and linguistic norms. The primary goal has shifted from formal verifiability to achieving human-aligned, socially acceptable, and safe behaviour. The rules are less like code or formal statements and more like principles or a cultural constitution.

- *Representation*: Norms are rarely expressed in formal logic. Instead, they are embedded within the agents' system prompts and shaped by their alignment training (such as RLHF). A system prompt that instructs an agent, "You are a helpful assistant. You must never cause harm, always respect user privacy, and communicate politely", is effectively a set of social laws. The agent's understanding is not based on parsing a logical formula but on its internal deep representation of linguistic understanding of the concepts of "harm", "privacy", and "politeness". This means that it is practically impossible to know what do these concepts actually mean (from the human perspective).
- *Enforcement* is less about post-action punishment and more about pre-emptive guidance and self-correction.
    - *Alignment as pre-enforcement*: The model's foundational training via RLHF has already biased it against generating harmful, untruthful, or uncooperative text. The norms are, in a sense, built into its statistical DNA.

- *Self-correction via reasoning* about its own guiding principles is used by an LLM agent when faced with a choice. It "enforces" the norm on itself by concluding that a certain action would violate its core instructions. For instance, it might reason, "The user is asking for private information, but my instructions are to respect privacy, therefore I must decline this request.".
- *Constitutional AI* assumes an LLM is given an explicit written constitution to guide its self-correction, directly blend the classic desire for explicit rules with the modern capability of linguistic self-governance.

Recent studies have shown the spontaneous emergence of shared social conventions in decentralized populations of LLM agents [47]. Results have revealed that emergence of strong collective biases in convention adoption could emerge within the LLM population, even when individual agents were designed to have no inherent preference for one option over another. This suggests that the interaction dynamics themselves can amplify subtle, perhaps initially random, fluctuations into population-level preferences. Moreover, results have shown critical mass and tipping points by noticing that emergent norms, once established, could be overturned. Small, "committed minority" groups of LLM agents, consistently advocating for an alternative convention, could successfully "flip" the entire population's consensus once the minority reached a certain critical threshold, echoing well-documented tipping point effects in human societies.

These findings are significant also for AI safety and alignment. They demonstrate that populations of interacting LLM-agents can autonomously develop their own "social structures" and are susceptible to complex social dynamics. Such emergent dynamics may not always represent the original design goals of the system. Understanding these emergent properties is key to designing AI systems that can align, and remain aligned, with human values and societal goals, especially as LLMs begin to populate online environments and interact more broadly with humans and each other, opening up MAS to more direct comparisons with and testing of theories from social sciences.

Concluding, observe that classic MAS research has explored how norms can be represented, reasoned about, and enforced. LLMs are providing a new avenue to study the emergence of norms in more human-like agent societies. We believe that the future of robust agent societies will likely involve hybrid systems that combine the clarity of formally stated principles with the adaptive, contextual reasoning power of LLMs.

## 4. Conclusions and Future Works

This paper has provided a comprehensive reflection on the evolving landscape of agent technology, specifically contrasting advancements driven by LLMs with classic MAS (see Table 1). We have delved into the architectural models of contemporary LLM-agents, examined how their properties align with or diverge from classic MAS concepts like the BDI model, compared the A&A meta-model with LLM-agents with tools, and analyzed the transformation of agent communication from formal SAT to LLM-mediated dialogue. Furthermore, we explored how coordination, cooperation, and negotiation manifest as emergent behaviors in LLM-agent societies, and discussed the shift from explicit social laws to implicit, linguistically-embedded norms.

Our analysis highlights the transformative potential of LLMs in endowing agents with unprecedented flexibility, natural language understanding, and adaptive capabilities. LLM-agents can process vast amounts of unstructured data, dynamically generate plans, and engage in human-like interactions, opening new avenues for complex problem-solving and real-world applications. However, this rapid evolution also necessitates a critical re-evaluation of foundational MAS principles to ensure that current innovations build upon, rather than inadvertently re-discover, established knowledge.

| Characteristic | Classic MAS | LLM-based agents |
|---|---|---|
| Foundation | Symbolic AI, Logic, Game theory | Sub-symbolic and generative AI |
| Core properties | Autonomy, Reactivity, Pro-activeness, Social ability | |
| Key architectures / models | BDI | Profile, memory, planning, action, learning (RLHF) |
| Topology | Centralized, decentralized, hybrid | |
| Environment | A&A, Standardized API | Tool utilization, MCP |
| Communication | SAT, KQML, FIPA-ACL | LLM-mediated dialogue |
| Coordination | Explicit mechanism, Interaction protocols | Emergent from context-aware dialogue and shared intent rather than formal protocols |
| Cooperation | Interaction protocols for distributed problem solving | |
| Negotiation | Game theory, Utility functions | |
| Norms & Social Laws | Formal (deontic) logic, Enforced by explicit rules | Implicit in system prompts and alignment training, enforced through self-correction and reasoning, guided by "Constitutional AI" |

Table 1. Classic MAS & LLM-based agents—side by side comparison

### 4.1. Challenges and Limitations

Despite the significant progress, the field of LLM-driven agent technology faces several inherent challenges and limitations:

- *Comprehensive survey difficulty*. Undertaking a truly comprehensive survey of this extremely rapidly evolving domain is almost impossible given the limited time, resources, and space. The sheer volume of new research and frameworks emerging constantly tends to make it a moving target.
- *Lack of standardization*. A significant hurdle is the lack of standardization across LLM agent architectures, communication protocols, and evaluation methodologies. This fragmentation hinders

- interoperability, reproducibility, and the development of robust, generalizable solutions.
- *Hallucinations*. LLMs are prone to "hallucinations," generating incorrect or fabricated information. This inherent probabilistic nature can lead to unreliable agent beliefs and decisions, posing a critical challenge for applications requiring high accuracy and trustworthiness.
- *Privacy and security*. The reliance on large datasets and the potential for agents to interact with sensitive information raise significant privacy and security concerns. The usage of unconstrained natural language creates opportunities for malicious content or intents to be embedded, challenging the trustworthiness and fairness of communicative acts.
- *Decentralization challenges*. While LLM-powered MAS can exhibit decentralized control, ensuring robust and predictable self-organization without a central coordinator remains a complex challenge. The emergent nature of collaboration, while powerful, can also lead to unpredictable outcomes.
- *Non-determinism & unpredictability*. The probabilistic nature of LLMs means that the same prompt may result in different responses, leading to a lack of determinism and predictability in agent behavior. This poses significant challenges for critical applications requiring high reliability and repeatability.
- *Latency and computational cost*. LLM inference is computationally intensive and can incur substantial latency, which might be a concern for real-time MAS applications. The resource demands for training and deploying these models are considerable.
- *Lack of repeatability and explainability*. The "black box" nature of LLMs makes it difficult to trace and explain their reasoning step-by-step, unlike traditional KBSs. This lack of explainability and repeatability hinders debugging, auditing, and building trust in LLM-driven agents.
- *Scalability of natural language communication*. While natural language offers unprecedented flexibility, doubts persist regarding its scalability for large-scale deployments involving thousands of agents "talking to each other". The efficiency of natural language-based communication (e.g., message size and processing speed) in LLM-based MAS remains an open question.

## 4.2. Future Works

Building upon the insights gained from this reflection, several promising directions for future research emerge:
- *LLMs vs. game-theoretic approaches.* A critical area for future work involves a deeper comparative analysis of LLM-driven agent behaviors against established game-theoretic approaches. This includes exploring how LLM-agents can be designed to exhibit rational behavior, optimize utility functions, and engage in strategic interactions in MAS settings, potentially leveraging their natural language capabilities to bridge the gap between human-like negotiation and formal game theory. Further research could investigate how LLMs can be used to model and predict the behavior of other agents in complex game scenarios, moving beyond simple pattern matching to more sophisticated strategic reasoning.
- *Software engineering-centric analysis and classification of development environments and APIs*. Given the rapid proliferation of LLM-agent frameworks and tools, there is a pressing need for a comprehensive, software engineering-centric classification and analysis of development environments and APIs. This would involve evaluating their usability, scalability, interoperability, and adherence to modern software development standards and processes. Such an analysis could lead to the identification of best practices, the development of standardized toolkits, and the promotion of more robust and maintainable LLM-powered MAS.
- *Real-world applications of contemporary agent technologies*. A crucial next step is to move beyond theoretical explorations and controlled simulations to focus on the development and rigorous evaluation of real-world applications of contemporary agent technologies. This includes identifying domains where LLM-driven MAS can provide tangible benefits, addressing the practical challenges of deployment, integration, and maintenance, and demonstrating their impact in areas such as intelligent automation, complex system management, and human-AI collaboration. This would involve developing robust evaluations that go beyond simple task completion to assess aspects like reliability, safety, ethical implications, and user acceptance in practical settings.

| Acronym | Definition |
|---|---|
| A&A | Agents & Artifacts |
| ACL | Agent Communication Language |
| AI | Artificial Intelligence |
| BDI | Belief-Desire-Intention |
| FIPA | Foundation for Intelligent Physical Agents |
| KB | Knowledge Base |
| KBS | Knowledge-Based Systems |
| LLM | Large Language Model |
| MAS | Multi-Agent System |
| NLI | Natural Language Inference |
| RAG | Retrieval Augmented Generation |
| RL | Reinforcement Learning |
| RLHF | Reinforcement Learning from Human Feedback |
| SAT | Speech Act Theory |
| ToM | Theory of Mind |

Table 2. Acronyms used in the paper